# Flat borophene films as anode materials for Mg, Na or Li-ion batteries with ultra high capacities: A first-principles study

Bohayra Mortazavi[*,1], Obaidur Rahaman [1], Said Ahzi[2], Timon Rabczuk[3]

[1]*Institute of Structural Mechanics, Bauhaus-Universität Weimar, Marienstr. 15, D-99423 Weimar, Germany.*

[2]*Qatar Environment and Energy Research Institute, Hamad Bin Khalifa University, Doha, Qatar*

[3]*College of Civil Engineering, Department of Geotechnical Engineering, Tongji University, Shanghai, China.*

## Abstract

Most recent exciting experimental advances introduced buckled and flat borophene nanomembranes as new members to the advancing family of two-dimensional (2D) materials. Borophene, is the boron atom analogue of graphene with interesting properties suitable for a wide variety of applications. In this investigation, we conducted extensive first-principles density functional theory simulations to explore the application of four different flat borophene films as anode materials for Al, Mg, Na or Li-ion batteries. In our modelling, first the strongest binding sites were predicted and next we gradually increased the adatoms coverage until the maximum capacity was reached. Bader charge analysis was employed to evaluate the charge transfer between the adatoms and the borophene films. Nudged elastic band method was also utilized to probe the ions diffusions. We calculated the average atom adsorption energies and open-circuit voltage profiles as a function of adatoms coverage. Our findings propose the flat borophene films as electrically conductive and thermally stable anode materials with ultra high capacities of 2480 mAh/g, 1640 mAh/g and 2040 mAh/g for Mg, Na or Li-ion batteries, respectively, which distinctly outperform not only the buckled borophene but also all other 2D materials. Our study may provide useful viewpoint with respect to the possible application of flat borophene films for the design of high capacity and light weight advanced rechargeable ion batteries.

Keywords: *flat borophene; anode material; first-principles; 2D material; high capacity*

*Corresponding author (Bohayra Mortazavi):  bohayra.mortazavi@gmail.com;

## 1. Introduction

While the demand for more efficient and high capacity lithium-ion batteries continues to rise, two-dimensional (2D) materials have recently emerged as promising electrode materials for such batteries. This is primarily due to their ability of maximal metal ions adsorption as well as fast ion diffusion along their fully exposed surfaces. The successful synthesis of graphene [1–3] a semi-metallic 2D material consisting of planar arrangement of carbon atoms with honeycomb lattice, followed by its widespread applications triggered the synthesis of other 2D materials like hexagonal boron-nitride [4] silicene [5] and transition metal dichalcogenides [6,7]. The potential for the integration and fabrication of heterostructures with tuneable properties make the 2D materials attractive for their application in various products such as rechargeable ion batteries [8–11]. Although carbon based 2D materials demonstrate unique and remarkable properties that are suitable for designing electrode materials, their capacities are rather limited. The commercially used graphite, although possessing many desirable qualities like cyclic durability, low manufacturing cost and high columbic efficiency, present a low theoretical specific capacity of 372 mAh/g [12]. As an alternative, application of bulk silicon with high theoretical capacity of 4200 mAh/g is prohibited because of the degradation due to drastic volume change [13,14]. On the other hand, the 2D materials that have many advantages as anode materials have moderately high capacities, for instance, $MoS_2$ [15–22], silicene [23] and germanene [24] have specific capacities of 912 mAh/g, 954 mAh/g and 369 mAh/g respectively. Thus, the search for a novel 2D material that is suitable to be utilized as an anode material but with a higher capacity is still ongoing.

In line with the current development in the synthesis of 2D materials, recently three different 2D boron sheets called borophene have been successfully synthesized by epitaxial growth of boron atoms on Ag surface under ultrahigh-vacuum conditions [25,26]. The existences, stability and electronic properties of these structures were already theoretically predicted before their experimental fabrication [27,28]. These new borophene structures, in accordance with the theoretical predictions, demonstrate metallic characters [25–27,29,30] and thus are good electric conductors, suitable as anode materials in rechargeable ion batteries. However, the real application of these 2D nanomembranes as an anode material in rechargeable ion batteries requires considerable amount of efforts. In this case, state of the art computational studies [31–35] can be useful only to predict some key parameters such as the charge capacity, voltage curves and diffusion barriers. However, with respect to the application of borophene films in batteries, there exist numerous technical issues that may prevent their practical usage. One should consider that the borophene films have been grown on the substrate and for their application in batteries they should be detached from the substrate, and be producible in large scale and low cost as well. Moreover, they may face stability issues in the in typical industrial battery processing conditions and they may also require specific electrolyte design.

A high capacity of an anode material primarily depends on strong adsorption of adatoms on it. Thus, we used density functional theory (DFT) first-principles calculations to study the adsorption of various metal atoms like Li, Na, Mg and Al on four different flat borophene films, two experimentally synthesized and the two others theoretically predicted. The adsorption energies and average open-circuit voltages are reported for each case. The theoretical specific capacities of each borophene sheet and for different adatom types were estimated, some of which demonstrate unprecedentedly high values. The diffusion of adatoms through the anode material is another crucial factor in deciding the rate performance of the battery. Therefore, we used nudged elastic band (NEB) method to assess and compare the diffusion of different adatoms over a flat borophene membrane. Mechanical and thermal stability of saturated borophene films was also verified by performing ab-initio molecular dynamics (AIMD) simulations. As compared with buckled borophene [25], our first-principles results suggest that the flat borophene films can be considered as more promising anode materials for rechargeable ion batteries. This conclusion was drawn due to the fact that flat borophene films can yield higher capacities in comparison with the theoretical estimations [31,34,36,37] for the buckled borophene. Moreover, we discuss that flat borophene structures react more stably upon the metal adatoms coverage in comparison with the buckled borophene. We note that flat borophene films were reported to attach weakly to the metal substrate [26] and such that the isolation of flat borophene films are probable. Nonetheless, in term of ions diffusion, flat borophene films cannot compete with buckled borophene with its extraordinary ultra-low diffusion barriers for metal-ions diffusions [31,36,37].

## 2. Computational method

Vienna ab initio simulation package (VASP) [38,39] in conjunction with the Perdew-Burke-Ernzerhof (PBE) generalized gradient approximation exchange-correlation functional [40] was used for the density functional theory (DFT) calculations performed in this study. We should note that boron bond has quite unique properties which may be significantly affected by the functional employed in the DFT calculations (see Ref. [41,42]). In this work, we considered the van der Waals interactions using the semiempirical correction of Grimme [43], as implemented in VASP. For reporting the binding energies of adatom adsorptions, this approach is a valid choice [44,45].

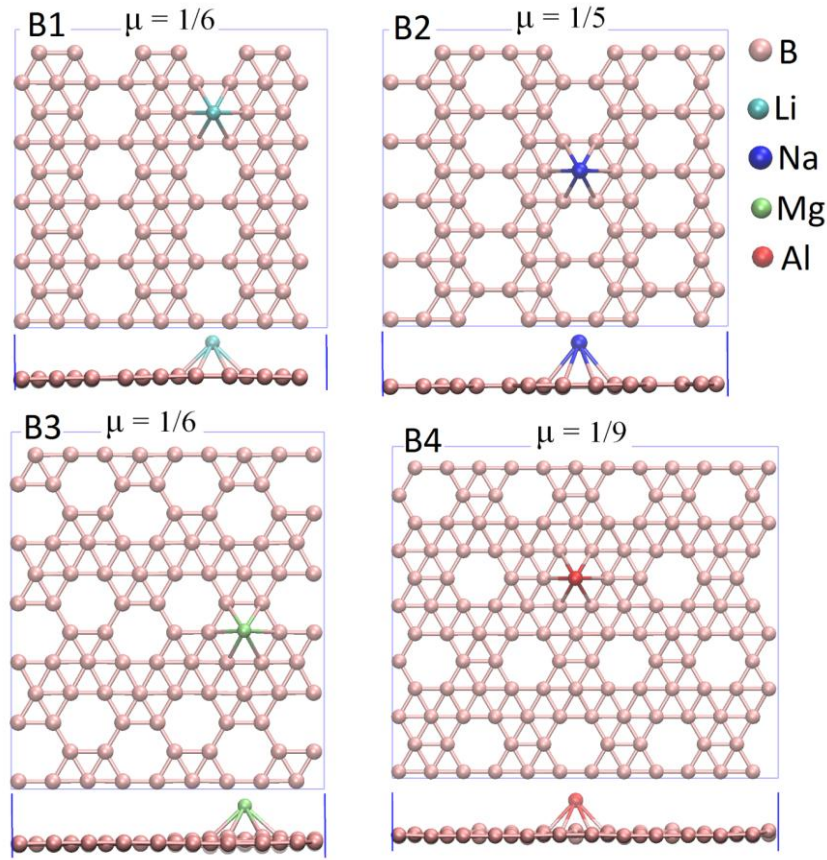

Fig-1, Top and side views of four different borophene sheets (B1, B2, B3 and B4) explored in this study. For each sheet, binding of an adatom at the most favourable adsorption site are illustrated. Here µ, represents the hexagon hole density which is defined as a number of hexagonal vacancies divided by the total number of atoms in the original triangular borophene lattice. VMD [46] software was used to illustrate these structures.

We studied the application of four different flat borophene films as anode materials for rechargeable ion batteries. For simplicity, we named them B1, B2, B3 and B4, which the first two were experimentally synthesized and the other two were theoretically predicted [28]. 2D boron membranes contain hexagonal and triangular lattices in which patterned missing of boron atoms create hexagon holes. In this way, the atomic structure of borophene films can be presented as a mix of hexagonal vacancies and triangular units. The exact ratio between hexagonal and triangular units is sensitive to several factors such as the synthesis procedure, substrate material and other experimental factors [25,26,28]. Such that various borophene polymorphs can be introduced by a "hexagon hole density" (µ), which is defined as a number of hexagonal vacancies divided by the total number of atoms in the original triangular lattice. In this way, for the buckled borophene with perfect triangular structure [25] the μ is equal to zero. In the present investigation, four supercell structures of borophene sheets B1, B2, B3 and B4 were constructed with 75, 80, 90 and 128 boron atoms, respectively, as they are illustrated in Fig. 1 along with their hexagon hole densities. Periodic boundary conditions were applied in all directions with 20 Å vacuum layer between the sheets, in order to avoid the image-image interactions [23,24]. At the first step, the most favourable binding sites for each adatom (Li, Na, Mg and Al) were identified on the different borophene sheets (B1, B2, B3 and B4). In this work, adatoms were randomly but uniformly positioned on both sides of the sheet in the previously defined favourable sites. The adatom coverage was increased randomly and in a step by step manner to simulate gradual adatom intercalation during a battery charging process. These initial structures were then treated with conjugate gradient method energy minimization. For this step, the projector augmented wave method [47] was employed with an energy cutoff of 400 eV with $10^{-4}$ eV criteria for energy convergence and 4×4×1 k-point Monkhorst Pack mesh [48] size for Brillouin zone sampling. After successful completions of energy minimization, single point calculations were performed to report the energies and also to obtain electronic density of states (DOS) of the systems in which the Brillouin zone was sampled using a 10×10×1 Monkhorst k-point mesh size and the energy cutoff was set at 500 eV. This was followed by Bader charge analysis [49] in order to estimate the charge transfer between the adatoms and the substrates and also to finally report the charge capacity of different borophene films for each considered ion. The adatom diffusion pathways over the borophene sheets and the corresponding energy barriers were calculated using nudged elastic band (NEB) [50] method. For the NEB calculations, we only considered the B1 borophene with 30 atoms. The reversibility of borophene structure deformations after maximum ion adsorption was also demonstrated using ab-initio molecular dynamics (AIMD) simulations. For the AIMD simulation in NPT ensemble, Langevin thermostat was used. The time step of AIMD calculations was set to 1 fs using the 2×2×1 k-point mesh size.

## 3. Results and discussions

Stronger, or in another word more negative adsorption energy of an adatom on the substrate is of vital importance for it to be suitable as an anode material of metal ion batteries. Thus, an in depth analysis of metal atoms (Li, Na, Mg and Al) adsorption on flat borophene sheets are carried out in this work. The identification of suitable adsorption sites are a prerequisite for the theoretical predictions of adsorption energies. To this end, we created several initial structures with metal ions positioned near all possible adsorption sites on the borophene sheets. These structures were then minimized using conjugate gradient methods. Fig. 1 shows the four borophene sheets B1, B2, B3 and B4 with various adatoms adsorbed on the site that produced the strongest binding as estimated by the adsorption energy, $E_{ad}$, defined by:

$$E_{ad} = E_{BM} - E_B - E_M \quad (1)$$

where $E_B$ and $E_{BM}$ are the total energies of borophene before and after metal atoms adsorption and $E_M$ is the per atom lattice energy of the metal adatoms (M =Li, Na, Mg or Al). Interestingly, for the all studied borophene films and different adatoms, we found that the hexagonal hollow sites present the maximum binding energy. For the B1, B3 and B4 borophene films, the second most favorable binding site was found to be on the top of the center atom in the fully occupied hexagonal lattice. For structure B2, all the trials for the search of second adsorption site resulted into the first binding site. We found that energy minimizations starting from any other initial positions resulted into the shift of the adsorbed ion towards either of the binding sites identified earlier. This indicates negligible energy barriers between the initial and final positions. The adsorption energies of the strongest binding are summarized in Table 1. The strongest adsorption energy of Li on B1 was found to be -1.779 eV which matches quite well with the value of -1.766 eV estimated by another DFT study [32]. The adsorption energy of the second most favorable site for Li was found to be -1.060 eV. The lowest adsorption energies of Li on B2, B3 and B4 were predicted to be -1.505 eV, -1.795 eV and -1.054 eV, respectively. The lowest adsorption energies of Al on all the borophene sheets were all very close to zero or positive. On the other hand the lowest adsorption energies of all the borophene sheets were negative for Li and Na indicating exothermic formation.

Table 1: Adsorption energies of Li, Na, Mg and Al atoms on B1, B2, B3 and B4 borophene sheets.

| | Adsorption Energy (eV) | | | |
|---|---|---|---|---|
| Structure | Li | Na | Mg | Al |
| B1 | -1.779 | -1.51 | -0.652 | -0.0622 |
| B2 | -1.505 | -1.318 | -0.199 | 0.48 |
| B3 | -1.795 | -1.53 | -0.696 | -0.144 |
| B4 | -1.054 | -0.786 | 0.517 | 1.05 |

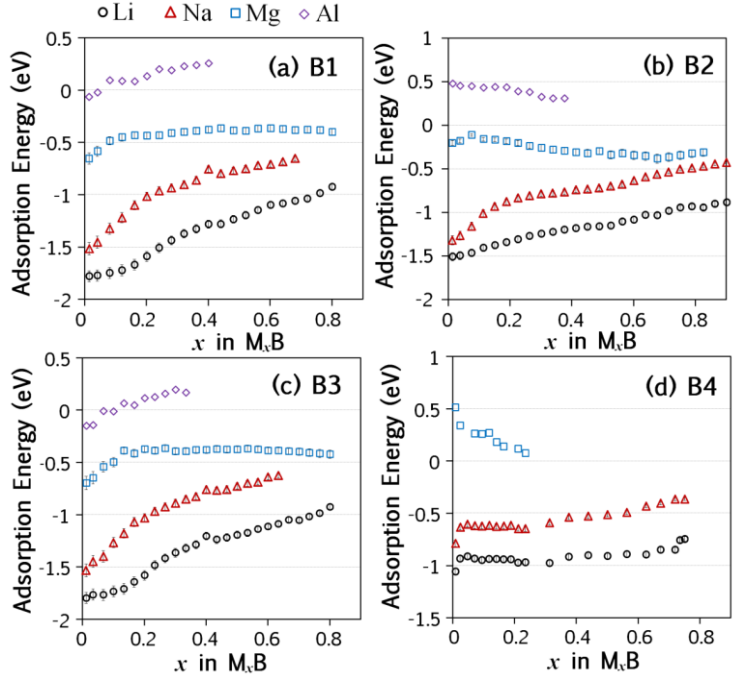

Fig-2, Average absorption energy for Li, Mg, Al or Na atoms on B1, B2, B3 and B4 flat borophene films as a function of ions coverage.

Since both the composition and structure of the anode material can drastically change with increasing coverage of adsorbed metal ion, it is very important to study this effect for the materials considered in this work. To this end, we carried out energy minimizations of borophene sheets with different coverage of metal adatoms, until no more metal adatoms could be adsorbed on the both sides of the borophene surfaces or the adatoms were not able to transfer their electrons to the borophene (maximum capacity is reached). In this work, we gradually increased the adatoms coverage, first by positioning the metal atoms randomly and uniformly on the both sides of the hexagonal hollow sites. After filling the all hexagonal hollow centers, we then positioned the adatoms on the second favorable site till the capacity is reached. Since no second lowest adsorption energy site was identified for B2, for this borophene film after filling the all hexagonal holes, the metal atoms were placed to be the center of the triangles, surrounded by three adatoms already filled the hexagonal hollow rings. After the achievement of energy minimization, single point energy calculations were carried out using the final structures. The average adsorption energies were calculated using the following equation:

$$E_{\text{av-ad}} = \frac{(E_{\text{MnB}} - n \times E_M - E_B)}{n} \qquad (2)$$

where $E_B$ is the total energy of borophene before adsorption, $E_{\text{MnB}}$ is the energy of borophene with n metal adatoms adsorbed on it. $E_M$ is the per atom lattice energy of the metal adatoms (M=Li, Na, Mg or Al). Fig. 2 shows the evolution of average adsorption energies with increasing coverage of various metal adatoms on the four borophene sheets. Except for the case of Mg atom, the average adsorption energy gradually becomes less negative with increasing metal atoms coverage, indicating weaker and weaker binding. This is primarily caused due to the enhanced repulsions between the positively charged metal adatoms. As indicated by the adsorption energies, throughout the whole range of x in $M_xB$, Li atoms adsorbed more strongly than the Na atoms and Na atoms adsorbed more strongly than the Mg atoms. For Al atoms, for all studied borophene films the average adsorption energies were positive for small coverages indicating that Al storage in these borophene sheets were endothermic. That makes these borophene sheets unsuitable for Al storage and thus we did not consider Al further in this study. For B1, the highest metal atoms adsorption, $x_{max}$ was found to be 0.76, 0.76 and 0.67 for Li, Mg and Na respectively. The slight variations of the $x_{max}$ values, for different borophene sheets, can be attributed to the structural details of the sheets. When more metal atoms were added to the system beyond the saturation points, they were pushed out of the borophene surface and formed a second layer on top of the first layer of adsorbed metal atoms.

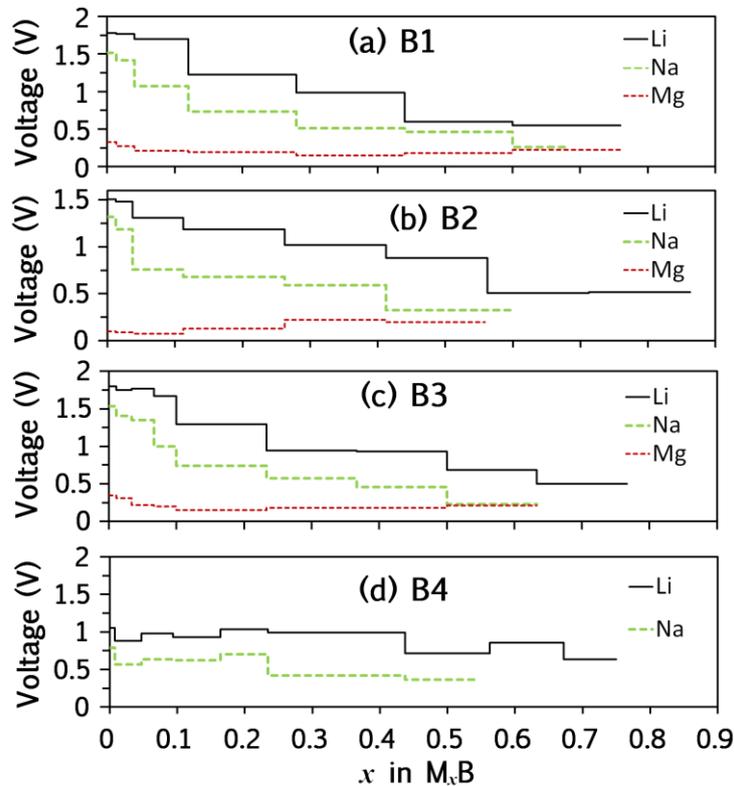

Fig-3, Calculated voltage profiles as a function of adatoms coverage for borophene nanomembranes. The voltage curves are plotted up to the coverage that yields the maximum capacity.

The open-circuit voltage profile is a crucial measure of the performance of an anode material. It can be estimated by measuring the voltage over a range of metal atoms coverage. The charge/discharge process of borophene relates to the common half reaction vs M/M+:

$$(x_2 - x_1)M^+ + (x_2 - x_1)e^- + M_{x_1}B \leftrightarrow M_{x_2}B \qquad (3)$$

Then the average voltage of $M_xB$ in the coverage range of $x_1 \leq x \leq x_2$ can be estimated by the following equation [51]:

$$V \approx \frac{(E_{M_{x_1}B} - E_{M_{x_2}B} + (x_2 - x_1)E_M)}{(x_2 - x_1)e} \qquad (4)$$

where $E_{M_{x_1}B}$, $E_{M_{x_2}B}$ and $E_M$ are the energies of $M_{x_1}B$, $M_{x_2}B$ and metalic M respectively. Here e is the charge of an adatom, for Na and Li is equal to one and for Mg is two. Ideally Gibbs Free Energy (G(x) = ΔE + PΔV - TΔS) should be considered for the calculation of average voltage. However, since PΔV and the entropic term TΔS are negligible compared to the energy term ΔE, only energy is considered in further calculations [51,52].

Fig. 3 shows the voltage profiles of the four borophene sheets for various metal adatoms. It is important to note that for the all the cases, the voltage remains positive during the whole range of coverage values. A negative value of voltage means that the metal ion prefers to form metallic states instead of adsorbing on the anode surface. Thus, the lack of negative voltage indicates that all these borophene sheets are suitable to be the anode materials for Li, Mg and Na-ion batteries, within the range of coverage explored in this work. Within the range of coverage values, the open circuit voltage for Li decreases from ~1.8 V to ~0.6 V for B1, from ~1.5 V to ~0.5 V for B2, from ~1.8 V to ~0.5 V for B3 and from ~1 V to ~0.6 V for B4 borophene. The variation of open-circuit voltage was relatively smaller across the range of coverage values for Mg. Evidently, higher open-circuit voltage is expected for Li-ion batteries than Na-ion batteries and the lowest was expected for Mg-ion batteries. These predicted range of voltages are desirable for anode materials.

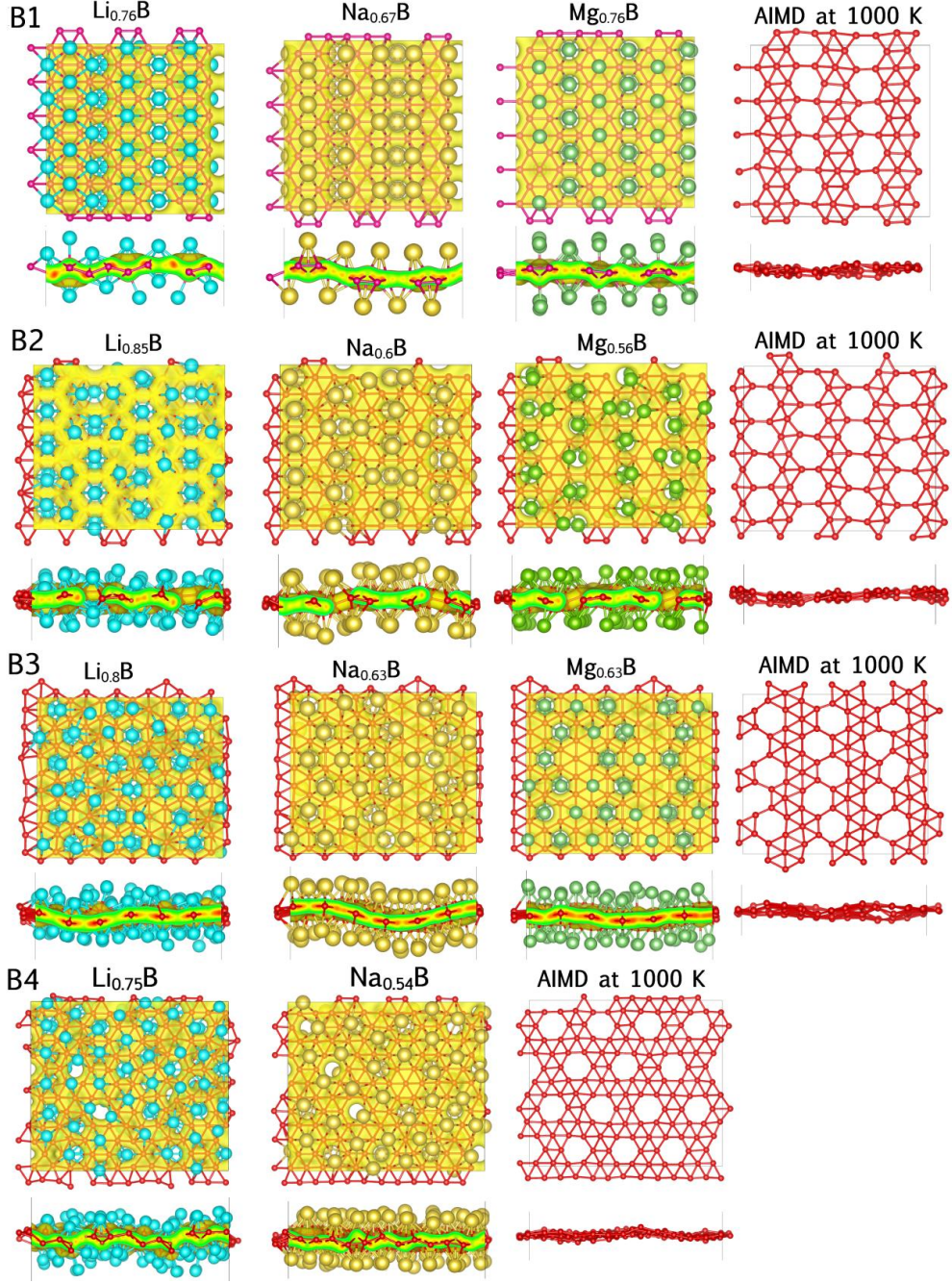

Fig- 4, Top and side views of energy minimized B1, B2, B3 and B4 borophene films with maximum capacity for Li, Na and Mg ions. According to our modeling results, borophene films are flexible and deform slightly upon the adatoms adsorption. After removing the intercalated ions, AIMD simulation at elevated temperatures well confirm the thermal and mechanical stability of borophene nanomembranes during charging or discharging cycles. VESTA [57] package is used to illustrate these structures along with their charge distributions.

In Fig. 4, we illustrate energy minimized borophene films with the maximum capacity for Li, Na and Mg atoms. Bader charge analysis [49] was employed to evaluate the charge transfer between the adatoms and the borophene films. The maximum charge capacity of each borophene for a particular adatom was predicted according to the coverage that either the further increasing of the ions coverage on the surface does not transfer additional electrons to the substrate or in which the further adsorbed atoms are pushed out of the borophene surface. Using the Bader charges analysis, the maximum charge capacities were acquired. For Li atoms storage, B1, B2, B3 and B4 borophene nanomembranes yields remarkably high capacities of around 1880 mAh/g, 2040 mAh/g, 1980 mAh/g and 1840 mAh/g, respectively. The Na atoms mainly due to their larger atomic sizes, however present the minimum but yet outstanding charge capacities of 1640 mAh/g, 1470 mAh/g, 1550 mAh/g and 1340 mAh/g over B1, B2, B3 and B4 single-layer borophene sheets. It is worthy to mention that based on the recent DFT study [32], the charge capacity of flat borophene films are predicted to be the same for Li and Na atoms. Their findings are in contradiction with our results for flat borophene films and also previous theoretical studies for buckled graphene [31,36,37] in which the capacity was found to be considerably higher for Li atoms storage as compared with Na atoms. Moreover, according to our results the capacity of flat borophene films for considered metal ions storage are very close whereas in the earlier investigation [32] the B1 film was suggested to yield much higher capacity as compared with the B2 film. Most fascinatingly, according to our modelling the Mg adatoms can yield ultra high capacities around 2400 mAh/g over flat borophene films. In this case, B1 borophene yields a charge capacity of around 2480 mAh/g which is distinctly the highest charge capacity accessible in comparison with other synthesized 2D materials. The Mg atoms capacity for B2 and B3 borophene sheets were predicted to be around 2400 mAh/g and 2330 mAh/g. However, for the this metal atom the B4 borophene presents a positive adsorption energy and such that it cannot be used as an anode material for Mg ions storage. In Table 2, we summarize the capacity of various borophene structures for different metal atoms storage. We rank the structures according to their hexagonal hole density, trying to establish a relation between the capacity and the hole density of borophene films. For Li ions, as a general trend it can be seen that by decreasing the hole density the capacity decreases. For Na atoms, the charge capacity is less dependent to the hole density. For the Mg ions storage, the flat borophene films were found to present close capacities which are distinctly higher than that of the buckled borophene. Since the rechargeable Li-ion batteries are currently the most popular energy storage systems, our study highlights that the higher capacities might be accessible through increasing the hole density in the flat borophene structures.

Table 2: Summary of charge capacity of buckled, B1, B2, B3 and B4 borophene sheets for various metal ions storage. The capacities of buckled borophene for different ions were extracted from our earlier study [36].

| Structure | Hexagon hole density | Charge capacity (mAh/g) | | |
|---|---|---|---|---|
| | | Li | Na | Mg |
| Buckled | 0 | 1720 | 1380 | 1960 |
| B4 | 1/9 | 1840 | 1340 | ----- |
| B1 | 1/6 | 1880 | 1640 | 2480 |
| B3 | 1/6 | 1980 | 1550 | 2330 |
| B2 | 1/5 | 2040 | 1480 | 2400 |

During the charging and discharging cycles of a rechargeable battery operation, depending on the current direction the ions coverage increases or decreases. Such a frequent processes consequently induce compositional and structural change of the anode material [36]. Regarding the intercalated flat borophene films with Li, Na or Mg atoms, our results illustrated in Fig. 4 suggest flexibility but slight deformation of the borophene membranes upon the adatoms adsorption. Such deformations are expectable due to strong interaction of ions with substrate. However for the application of a material in rechargeable ion batteries these lattice deformations should not lead to persistent structural changes such as bond breakages or defect formations. For the all saturated films, we accordingly removed the adatoms and then performed AIMD simulations at room temperature (300 K) and 1000 K. Our analysis of the AIMD trajectory results for 10 ps of simulations at both 300 K and 1000 K (as illustrated in Fig. 4) well confirm that after the removal of adatoms, even at elevated temperatures the borophene membranes are stable and the adatoms adsorption could not result in structural damages. We note that our recent theoretical investigation regarding the buckled borophene film reveals considerable structural deformations due to the Li, Na, Mg and Ca adatoms adsorptions [36].

For the all considered borophene sheets with different coverage of Li, Mg or Na atoms, we performed electronic density of states (DOS) calculation. Our calculated total DOS confirm that in the all cases, at the zero state energy (Fermi level) the DOS is not zero which consequently demonstrate metallic character. We remind that all predicted borophene are metallic [28]. Presenting electronic conductivity is a highly desirable factor for the suitability of a material as anode or cathode in rechargeable ion batteries. Internal electronic resistance and ohmic heats produced during the battery operation are directly proportional to the electronic conductivity of the cathode and anode materials. Good electronic conductivity of borophene films consequently enhances its application prospect for the design of advanced rechargeable ion batteries with high efficiencies.

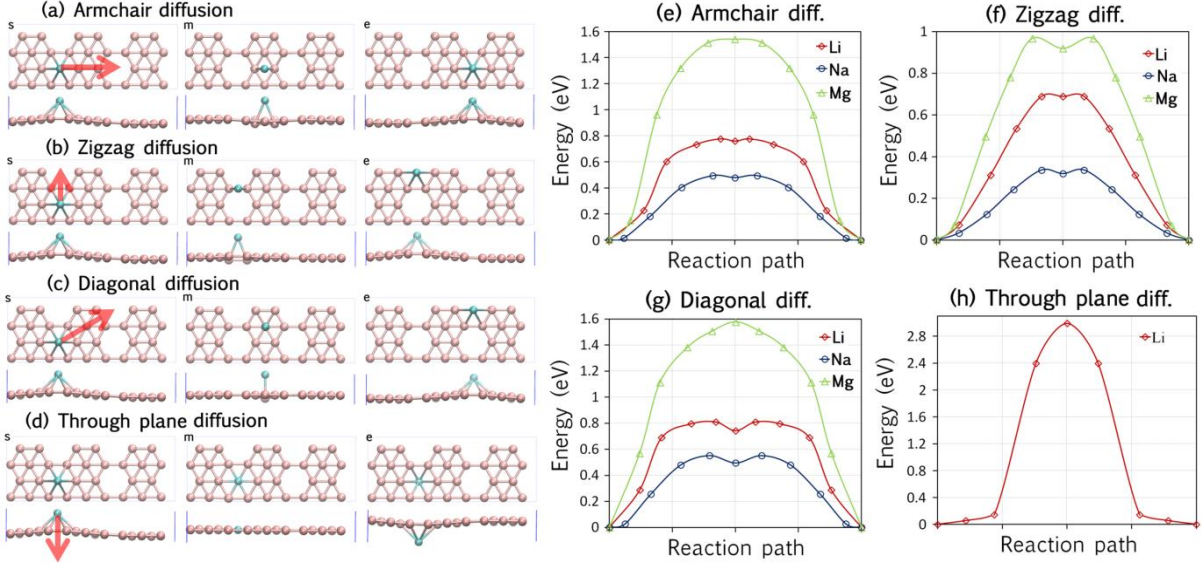

Fig-5, Top and side views of a single Li atom diffusion pathways (the arrow indicates the diffusion path) along (a) armchair, (b) zigzag, (c) diagonal and (d) through the plane, directions. For the considered diffusion directions, we illustrated the structures at three steps denoted with; "s", "m" and "e", indicating, starting, middle and end points, respectively. (e-h) Nudged-elastic band results for the energy barriers of Li, Mg and Na adatom diffusions along different directions for the B1 borophene. According to our DFT results, Mg or Na ions cannot pass through the hexagonal holes in the flat borophene lattices.

The charge-discharge rates of rechargeable batteries (commonly referred to C rate), which is a key factor in determining the performance of electrodes, depends on the diffusion barriers of metal adatoms through the anode materials. A low diffusion barrier and high mobility of the metal atoms are highly desirable. In order to investigate the optimal diffusion paths of metal-ions on the borophene surface and estimate the corresponding diffusion barriers, we employed nudged elastic band (NEB) method. Fig. 5 shows the top and side views of several possible pathways of diffusion of Li, Na or Mg atom over B1 borophene sheet. As illustrated, four possible pathways were explored, along the armchair, zigzag, diagonal and finally penetrating through the thickness of the sheet. The corresponding energy barriers for each pathways and for various ions are compared in Fig. 5 (e-h). Comparison of the diffusion energy barriers suggest that the diffusion is more favourable along the zigzag direction through hopping two the neighbouring hexagonal hole, as indicated by the lowest energy barrier. The diffusion energy barriers over the B1 borophene surface for Li, Na and Mg ions are therefore predicted to be 0.69 eV, 0.34 eV and 0.97 eV. Our predictions for Na and Li adatoms energy barriers are in close agreement with estimated values of 0.33 eV and 0.66 eV reported by the earlier DFT study [32]. According to our NEB modeling results, Mg or Na ions cannot pass through the tight hexagonal holes in the borophene lattices. We next compare the diffusion barrier of Li adatom on flat borophene and on other 2D materials. The diffusion barrier of Li on most Mxenes is 0.05–0.15 eV [53,54], on phosphorene is 0.13–0.76 eV [55], on graphene is ~0.37 eV [56], on silicene is 0.23 eV [23] and on germanene or stanene is 0.25 eV [24]. We remind that the Li diffusion energy barrier on the buckled borophene [36] is one order of magnitude smaller than our predictions for the flat borophene films. Although near the higher end as compared to other 2D sheets, the diffusion energy barriers of Li, Mg and Na adatoms on the flat borophene films estimated in this work are yet acceptable for an anode material for rechargeable ion batteries.

## 4. Conclusions

In summary, first-principles calculations were performed to probe the potential application of flat borophene films for Li, Mg and Na ion storage. Four different borophene structures were considered for this investigation. The results of our theoretical work strongly suggest that borophene sheets possess properties that are highly desirable for their applications as anode materials in rechargeable ion batteries. A single atom, irrespective of the atom-type, is predicted to adsorb preferably above the hexagonal hollow sites of the borophene lattice. The investigation on the effect of adatoms coverage on average adsorption energy suggests a gradual decrease in the adsorption energy with increasing adatoms coverage. Nevertheless, favorable exothermic adsorptions were predicted at high metal to substrate atomic ratio, irrespective of atom type. Amazingly, an ultra high capacity of around 2480 mAh/g was predicted to be accessible through the use of a flat borophene for Mg atoms storage. The storage capacities of Na and Li were also predicted to be as high as 1640 mAh/g and 2040 mAh/g, respectively. These specific charge capacities are notably higher than the values of other 2D materials, either experimentally observed or theoretically predicted. We also tried to establish a relation between the capacity and the hexagonal hole density of different borophene films. In this regard, for the Li-ion batteries our study highlights that the higher capacities might be accessible through increasing the hole density in the flat borophene structures. The average open-circuit voltages estimated for the battery application suggest a range from ~0.15 V to ~0.96 V for the different metal adatoms. This voltage range is desirable for low voltage batteries. Acceptable diffusion energy barriers of metal ions on borophene surfaces predicted by this work ensure reasonable charge-discharge rates of the batteries. After the removal of the adsorbed ions, AIMD simulations performed on the borophene sheets indicates the flexibility and remarkable thermal stability of the flat borophene nanomembranes. AIMD results also well confirm that metal atoms adsorption induce no persistent structural damages to the flat borophene lattices. Electronic density of states results for the all considered borophene sheets with different coverage of Li, Mg or Na atoms reveal metallic character meaning the good electronic conductivity which is a highly desirable factor for the efficient performance of rechargeable ion batteries. Our first-principles modeling results suggest the flat borophene nanomembranes as outstanding candidates for the use as anode materials because of their uniquely high charge capacities, good electronic conductivities, considerable adsorption energies, high thermal stability, acceptable diffusion rates and suitable open-circuit voltage profiles.

## Acknowledgment

BM, OR and TR greatly acknowledge the financial support by European Research Council for COMBAT project (Grant number 615132).